**Does the Great Firewall really isolate the Chinese? Integrating Access Blockage with Cultural Factors to Explain Web User Behavior**[1]

RUNNING HEAD: Internet Blockage and User Behavior

Harsh Taneja, Missouri School of Journalism, University of Missouri, Columbia, Missouri, USA

Angela Xiao Wu, School of Journalism and Communication, Chinese University of Hong Kong, Hong Kong[2]

**Contact Information:**

Harsh Taneja
Missouri School of Journalism
University of Missouri
181C Gannett Hall
Columbia, MO
65211-1200
USA

Harsh Taneja <harsht@u.northwestern.edu>

Angela Xiao Wu <angela.xiao.wu@u.northwestern.edu>

---

[1] We thank James Ettema for his helpful comments on earlier versions of this manuscript. The current version has also benefited greatly from extended discussion with James Webster, Stephanie Edgerly and Edward Malthouse. We are very grateful for the detailed and constructive feedback from the anonymous reviewers. Finally, we would like to acknowledge the terrific editorial support and direction from the editor of *The Information Society*. We wish to thank Rufus Weston who enabled the project by providing us access to the comScore data through BBC Global News. This research was partially supported by a Graduate Research Grant from Northwestern University.

[2] Both authors have contributed equally to this manuscript.

**Abstract**

The dominant understanding of Internet censorship posits that blocking access to foreign-based websites creates isolated communities of Internet users. We question this discourse for its assumption that if given access people would use all websites. We develop a conceptual framework that integrates access blockage with social structures to explain web users' choices, and argue that users visit websites they find culturally proximate and access blockage matters only when such sites are blocked. We examine the case of China, where online blockage is notoriously comprehensive, and compare Chinese web usage patterns with those elsewhere. Analyzing audience traffic among the 1000 most visited websites, we find that websites cluster according to language and geography. Chinese websites constitute one cluster, which resembles other such geo-linguistic clusters in terms of both its composition and degree of isolation. Our sociological investigation reveals a greater role of cultural proximity than access blockage in explaining online behaviors.

"A new information curtain is descending across much of the world," announced the U.S. secretary of state Hillary Clinton in her speech on "Internet Freedom," two months after attending the celebration to commemorate the fall of the Berlin Wall (Clinton, quoted in Cramer 2013, 1077). Her statement illustrates the common understanding that state blockage creates isolated communities on the World Wide Web. Such claims are based on an assumption that Internet users would use all websites if given access. On the contrary, a large body of research on global cultural consumption shows that audiences prefer products that are closer to their culture, even when they have access to products from abroad. Hence, we advocate a framework that integrates access blockage with other social structures, such as language and geography, to explain web user choices. This framework treats online state censorship as a form of cultural protectionism.

To test this conceptual framework, we focus on China where Internet censorship is most developed and has garnered enormous attention internationally. We examine global patterns of web usage to compare Chinese web usage with other regions not subject to state-imposed access blockage. First, we analyze traffic among the 1000 most visited websites globally, and discover that websites cluster according to geography and language into many culturally defined markets. The Chinese cluster is one such market, and we do not find it to be particularly isolated compared with other geo-linguistic clusters that take shape on the "open" Internets. Further, we examine the specific composition of the Chinese culturally defined market, and find that blockage is a rather limited explanation for it. Finally, we speculate how Chinese Internet users' behavior would change if the Great Firewall of China is lifted.

### Internet Censorship and the Internet in China

China's very first full-function linkage to the World Wide Web was realized in 1994, through a 64K international dedicated circuit provided by the Sprint Corp. of the United States. The Chinese government began to regulate online content and services in 2000, when Internet access was still confined to the elite strata in research institutes, state organs, and foreign companies (below 2% penetration rate). While strengthening ideological control over this new media, the state has also been taking an aggressive role in expanding internet infrastructures and information economy (Harwit and Clark 2001; Zhang 2006). Undergoing exponential growth, the size of Chinese online population has reached 564 million and the number of domestic websites 2.68 million by the end of 2012 (CNNIC n.d.). This combination of factors has resulted in China having the largest online population among all countries, a large and prosperous domestic Internet industry, and the most technically sophisticated Internet censorship.

State censors act upon all types of websites that they consider either violate social and moral norms (eg., sexually explicit content, see Jacobs 2012), or threaten the ruling power (Faris and Villeneuve 2008; Bamman, Brendan, and Smith 2012). Censorship is not restricted to news and political sites as users in nondemocratic regimes tend to engage with the seemingly "nonpolitical" realm in politically consequential ways. For instance, in China's personal blogosphere, social media, and video sharing sites, users develop distinct styles to express political sarcasm and initiate civic discourse as spinoffs of entertainment content. They also mobilize political protests using social networks established and maintained through leisure-oriented activities online, which makes any website with networking capacities potentially a target of censorship (MacKinnon 2008; Marolt 2011; Meng 2011; Xiao 2011; Yang 2009).

Chinese Internet censorship includes two types of measures: content censorship over domestic websites, and access blockage targeting websites outside the state's jurisdiction. Regarding the first measure, content censorship, it needs to be noted that the domestic online landscape is not at all monolithic in terms of ideology and information. To cope with the tremendous scale of online content production, the state relies increasingly on private corporations to monitor their own turfs, leading to a domestic information regime under a largely decentralized and heterogeneous discipline (MacKinnon 2009). Complex dynamics and vibrant user activities just discussed arise from this institutional arrangement – although censors watch over a wide range of online venues, the users are creative at bypassing the radar. Moreover, the state actively fosters a freer domain where people can vent frustrations and have fun, and where it can observe public opinions and adjust local policies (Hassid 2012; MacKinnon 2011; Zheng 2007). A recent study shows that censors of micro-blogs primarily expunge comments that "represents, reinforce, and spur social mobilization," but are more likely to permit vocal discontent with the government (King, Pan, and Roberts, forthcoming).

Despite all these ongoing contestations around content censorship in domestic cyberspace, public attention is overtly focused on the second measure of state censorship, i.e., access blockage, which is also our focus in this study. First coined overseas, this gigantic filtering mechanism is now widely termed as "the Great Firewall of China," or the GFW (Barmé and Ye 1997). The GFW allows regulators to prevent China-based Internet users to access targeted foreign websites through layers of technical intervention (Feng and Guo, forthcoming).

## "Free Flow of Information," Access Blockage, and User Preference

In the US policy discourse on international communications, the GFW is often equated with a digital version of the Iron Curtain that curbs the "free flow of information" (Cannici 2009; Tsui 2008). For example, as quoted in the opening, Hilary Clinton cautioned against "a new information curtain" that creates a vital "partition" on the global Internet (Clinton, quoted in MacKinnon 2011, 32). Terms such as "Chinternet," which spread widely since Google quit China over censorship dispute in 2010, clearly communicate the perception of the Chinese Internet as a distinct entity from the WWW as it is fenced in by a repressive regime (Chao and Worthen 2010; also see Mueller 2011).

These policy and popular discourses often presume that the GFW stops the Chinese from joining the world, embracing values of liberty and democracy, and thereafter pushing for democratization (Damm 2007; Tsui 2008). Sharing such beliefs, scholarly research documents in detail how the Chinese struggle to breach the GFW (Harwit and Clark 2001; Mulvenon and Chase, 2005; Xiao 2011). Inherent in this understanding are assumptions about Chinese users' Internet browsing choices, namely the websites they would choose if they were free to access what they wanted. In other words, the GFW is seen as preventing the Chinese from accessing foreign websites.

Resonating with the Cold War metaphors used to describe Chinese Internet filtering, a study on how communism's collapse affected international idea flows in Eastern Europe is particularly relevant. By measuring book translations before and after the collapse, it points out that the existing emphasis on state censorship in explaining low inflows of Western information obscures the role played by audience preferences (Abramitzky and Sin 2010). In this vein, to what extent does the GFW impair agentive information retrieval on the part of Chinese Internet users?

In the Global Internet User Survey (Internet Society 2012), compared with individuals from 19 other countries, the Chinese are not exceptional in their perceptions of Internet control-related issues; moreover, they tend to see more positive consequences of government regulations of the Internet. In fact, various surveys have found that in China tolerance of Internet censorship runs relatively high (Guo and Feng 2011; Li 2009). While censorship endorsement may result from certain cultural and historical factors, the fact that 89% (global average: 85%) of Chinese Internet users agree or strongly agree that they "have full access to all of the information that is available on the Internet" (Internet Society 2012, 20) suggest that the coercive impact of the GFW upon Internet use patterns is overestimated. As some ethnographic research suggests, many Chinese people are either unaware of the GFW, or unconcerned by it (Damm 2007, 282-285; F. Liu 2010). Therefore, the degree to which access blockage moderates the Chinese' ability to enact their preferences on the Internet requires large-scale empirical investigation, particularly as more and more media products become increasingly available globally.

Other than its presumptions about user preferences, we are also skeptical about the obsession over state-imposed blockage in global Internet governance because we see it as stemming from the ideological underpinnings of "free flow doctrine" initiated during the Cold War era (Nordenstreng 2011; Tsui 2008). Recent critical inquiries point to a problematic assumption of this discourse – only political domination hinders the free access to information. Economic interests that prevail in formally democratic countries are also at work (Cramer 2013).

In the same vein, we see state-imposed blockage like the GFW as infringing on people's freedom regardless of whether *coercive* prevention takes place. Internet governance should result from people's democratic decision-making, not from the arbitrary power of an authoritarian government. What we are against, fundamentally, is the Chinese users' condition of political subjection signaled by the operation of the GFW, rather than the specific type of interference it may have in people's browsing practices (see Skinner 1988).

In the following section we review related literature on global media consumption, and posit how access blockage combined with other cultural factors to explain Chinese web use patterns alongside the global ones.

**Global Internet Usage and Culturally Defined Markets**

The World Wide Web is conceived as an inherently global mass medium. Any user with access to an Internet enabled device, can potentially access the WWW regardless of her location. Therefore, any attempt at restricting access to content on the WWW is seen as preventing users from accessing information. Such a claim assumes that people are equally predisposed towards all content available on the WWW. Existing empirical studies suggest otherwise.

For instance, analyzing structure of country to country hyperlinks between websites, Barnett, Chung and Park (2011) find that sites in the same languages or those that focus on the same country tend to have more hyperlinks with one another than with sites in different languages or focusing on other countries. In a large scale analysis of Wikipedia, Hecht and Gergle (2010) found that each language Wikipedia differs from others in the concepts they cover as well as in the content of the common concepts covered. Likewise, Twitter ties form between people when they are in close geographic proximity, share the same language or when there is direct air connectivity between their locations (Takhteyev, Gruzd, and Wellman 2012). These findings indicate web users' tendency to show a predisposition towards local content.

In fact, not just web users but audiences across media, are predisposed towards local products. In general people prefer content that is in their own language, enjoy discussing it with their friends, and find it easier to derive meanings from local products than foreign products (Pool 1977). This tendency of audiences to prefer content that is closer to their culture is termed as cultural proximity. We see cultural proximity in play when global media consumption in the aggregate manifests as many culturally defined markets (henceforth CDM) (Straubhaar 2007). Each CDM is a media market where the content and the audience share common cultural traits, such as shared language and geographic proximity.

Language is one of the most obvious explanations for the formation of culturally defined markets (Straubhaar 2003) in part because large segments of audiences globally consume content mostly in languages they understand. Examples of such markets at the regional or transnational level include the Spanish speaking countries of Central and South America, mainland China and Hong Kong, Hindi/Urdu speaking (North) India and Pakistan, as well as much of the Arabic speaking Middle East (Curtin 2003; Straubhaar 2007). The different linguistic regions within Switzerland (German, French and Italian), India (Tamil, Telugu etc.) and Spanish speaking Hispanic audiences within United States (Ksiazek and Webster 2008) are examples of such markets within nation states. In other cases, CDMs are not geographically contiguous, e.g. the Portuguese speaking world (Brazil, Portugal, Mozambique, and other countries) or the English speaking world (US, Canada, UK, Australia, New Zealand, and other countries).

Other than language, the literature on cultural proximity also identifies geographical proximity between the place of production of media content and the audience under consideration as a factor. With abundant national and regional content available to audiences around the world, studies show evidence of cultural proximity in Asian countries such as India (Neyazi 2010) and Korea (Jin 2007), in Latin American countries such as Brazil and Ecuador (Davis 2003), and in Middle Eastern countries such as Lebanon (Kraidy 1999). Not all countries have an adequate market size and/or the wealth to produce copious amounts of national programming. Under such circumstances, the thesis of cultural proximity posits that audiences will gravitate towards content from nearby regions (Straubhaar 2007).

While language and geography are often correlated, since geographically contiguous regions tend to speak the same languages, CDMs also form when such regions do not share a common language. For instance Bollywood movies are quite popular in Afghanistan and much of the Arab World, even though Hindi (the language of most Bollywood movies) is neither spoken, nor understood in either of these countries. Another case in point is the Iberian Peninsula comprising Spain and Portugal. In such cases these territories share common histories of migration of ethnic groups, colonialism and cultural contact (due to geographical proximity, trade, etc.) (Straubhaar 2007).

Growth of CDMs is supported by policies of cultural protectionism practiced by many states, to preserve national cultures from foreign influence. Such policies largely involve two practices. First, many nation states impose quantitative restrictions on import of foreign content, such as setting quotas on number of foreign films or barring foreign television broadcasts during prime time. Second, states try to promote national production by offering subsidies and tax breaks to local culture industries (Baughn and Buchanan 2007). Initially protectionism did contribute to the growth of domestic cultural industries. However, in the last few decades, under the influence of liberalized international trade regimes many states have lifted protectionist measures on cultural products (Burri 2012). Yet, media audiences continue to be more inclined towards culturally proximate products.

Linking this well-established literature on cultural proximity in global media consumption to our current topic, we argue that people can enact their preferences much more easily towards culturally proximate content on the web. First, no regulatory clearances are needed to create Web pages in most countries, unlike television channels that require a broadcasting license or films that require clearances from censor boards or rating agencies. Second, the web provides many 'platforms' that allow ordinary users to share the content they have created, often with minimal resources, and upload it online. Hence it is possible to create, host, populate and popularize websites with minimum costs, if one has access to a computer connected to the Internet. Further, online content is usually free to consume for users browsing from any location unless the owner desires to restrict its access based on visitors' specific territories, or certain countries block specific websites from domestic viewing, such as in the case of China.

## Internet Access Blockage and User Behavior

Assuming that user behavior is driven by cultural proximity, we consider the impact of state censorship on online behavior to be analogous to that of cultural protectionism in traditional media. Both facilitate the formation of culturally defined markets. They achieve their effect in three ways. First, by definition, just as cultural import controls aim to restrict audience's media diets to national products or imports allowed by the state, Internet access blockage too constrains users from browsing foreign websites and thus confines them within the domestic Internet landscape.

More noteworthy, however, are the two indirect mechanisms by which the GFW accentuates the consumption of domestic websites. First, by blocking foreign sites, it acts like a trade barrier that enables growth of the domestic industry. The absence of foreign competition has clearly contributed to the dramatic success of the Chinese versions of Facebook, Twitter, and YouTube, as well as the fast expansion of Chinese online service providers in general (C. Liu 2010). The Chinese users, as a result, are left to choose from a wide range of domestically available products. Second, state censorship also limits people's knowledge about their media choices. Therefore, through altering the information milieu regarding media availability, the GFW may have shaped people's browsing preferences without them noticing it (Webster 2011). Therefore, it is quite likely that GFW has facilitated the consolidation of a Chinese CDM by enabling a large and varied supply of local products, as well as habituating people's preferences towards local products.

We expect people on average to keep consuming local websites, even if the GFW is lifted, making accessible erstwhile blocked foreign sites. Domestic producers "have a distinct advantage in the competition for an audience" once they manage to re-appropriate popular cultural formats from abroad (Pool 1977: 143). Thus after a point when local content becomes widely available, it trumps foreign content for people's attention. In addition, if the GFW does function largely through indirect mechanisms, as just discussed, Chinese are likely to continue their existing browsing patterns out of habit (Figure 1).

To summarize, rather than assuming that people use indiscriminately all the websites they have access to, we believe that they mainly prefer culturally proximate websites on the global Internet. In this view, access blockage is coercive only when it prevents users to browse culturally proximate content. Moreover, by potentially assisting the growth of domestic content (both in Chinese and China-focused) and limiting people's knowledge of foreign options, Chinese Internet censorship may have contributed to the formation of a CDM on the WWW. We further argue that, after a decade-long imposition, even if blockage is lifted today, Chinese users are unlikely to change their browsing behavior significantly (Figure 1). Any empirical exercise to explain Chinese web usage must incorporate cultural factors such as language and geographical focus of content alongside state-imposed access blockage, as we do in this study.

**Figure 1 about here**

## Data, Measures, and Methods

**Audience Duplication**

Most media users today have at their disposal a large number of media options. Yet they do not consume all these options and focus on a small number of such subsets commonly termed "media repertoires" (Taneja, Webster, Malthouse, and Ksiazek 2012). These repertoires could be based on a number of factors such as users' preferences and the platforms and content they are able to access. When analyzed in the aggregate, the composition of these repertoires indicates what media and/or content are consumed by the same audiences. Hence, in order to identify these repertoires we not only need to know how many people consume each media outlet, but how they move between different outlets. This information is often provided by a measure called *audience duplication*.

Audience duplication is the extent to which two media outlets are consumed by the same audiences. For instance, on any given day, if out of 100 people in a population, 20 people watched both FOX News and CNN, the audience duplication between them would be 20% for that day. Likewise, duplication can be calculated for all possible pairs of media outlets for a given audience. This results in a symmetric audience duplication matrix, where the elements $A_{i,j}$ represent the extent to which media outlets i and j have audiences in common. Such a matrix can be analyzed further to identify clusters of media outlets that have audiences in common. The earliest application of audience duplication was in identifying user defined program types, the subsets of television programs that were watched by the same set of users (Webster 1985). Webster and Ksiazek (2012) have recently used audience duplication to study the repertoires of websites commonly used together by US Internet users. In this study, we use such an approach.

We obtained audience duplication figures between all possible pairs of most popular websites. In doing so, we define duplication between any two websites as the % of unique users that visited both websites across any possible pair. This reveals the extent to which audiences move between all pairs of websites irrespective of the languages and geographical focus (if any) of these websites The resulting audience duplication matrix can help investigate, whether WWW consumption clusters into culturally defined markets. Before detailing on those methods we describe our source for audience duplication data.

**Data: comScore**

In this study, we use data from comScore[1], a panel based service that provides Internet audience measurement data once a month. It is currently the largest continuously measured audience panel of its kind. With approximately 2 million consumers worldwide in 170 countries under continuous measurement, the comScore panel utilizes a meter that captures behavioral information through a panelist's computer. Data are collected from both work and home computers of the panel members. Complementing the panel is a census-level data collection method, which allows for the integration of the aggregate level Internet behavior obtained through servers with audience information gained through the comScore panel.

ComScore organizes websites by Web domains and subdomains. We decided our sample to be the top 1000 Web domains (ranked by monthly unique users) in the world, as this number not only captures most of the domains that 99% of Web users visit, but ensures an adequate representation of sites in different languages and different geographies. For many large websites such as Google, the different geo-linguistic variants are classified as separate domains (e.g., www.google.es , www.google.de etc.).  For certain large domains such as Wikipedia, language versions are sub-domains of the main domain (e.g., es.wikipedia.org). In such cases, these sub-domains have been considered in the final sample instead of restricting to top level domains. These data reflect traffic during June 2012 and 973 websites were included in the final sample. These covered 50 languages in all (many sites were in multiple languages). For each one of the 973 websites, we obtained its audience duplication with all other 972 sites. Thus the final dataset has 472878 ((973 *972)/2) pairs of audience duplication numbers.

We collected our censorship data in November 2012 from GreatFire.org, an online organization that shares real-time and historical information about GFW blockage since early 2011.[2] GreatFire allows users to access blockage records of thousands of websites it routinely monitor, as well as to test their own URLs. Given the ever adjusting nature of GFW, for the majority of our sample websites that have established GreatFire profiles (several tests per month), we code censorship data based on whether they had been blocked in the past 30 days; for the remaining fraction of websites not monitored by GreatFire, we tested them on the spot and noted the results (either blocked or not). A total of 99 sites were blocked across genres including 21 entertainment sites, 18 file and photo sharing services, 13 social media sites and 4 news sites and 2 portals.

One of the authors visited each website and noted all languages each site offered content in. Additionally we relied on the self-description, metdata and third party sources (such as Alexa.com) to ascertain the geographic focus of the website. For about one third of the sites, where we could not assign a country we classify them as Global. Most of the sites that fall into the "global" category have subdomains that vary by language and at times have tailored content for different countries.

**Methods**

We used a number of analytical procedures on the audience duplication data. First, we analyzed the resulting audience duplication matrix using a set of network analytic tools. These suggested high evidence of underlying clusters of websites with highly duplicated audiences. Second, we conducted a hierarchical clustering of this matrix to identify these repertoires of websites commonly consumed together. These repertoires seemed well aligned to global Internet usage patterned by the presence of CDMs. We found that a large number of websites focusing on mainland China constitute one such cluster. We examined the average distance of each of these clusters to the rest of the network to understand their relative isolation. Further, we examine the membership of websites to the Chinese cluster based on their level of cultural proximity. In the section that follows, where we report our findings, we also describe these analytical procedures in greater detail.

## Analysis and Results

### Analyzing Global Online Audience Flows

As already stated, we conceptualized the audience duplication matrix as a network with the 973 websites as the nodes and % of unique users of total Internet users who visit both websites as the ties. However, since any two websites are expected to have some amount of audience overlap, we considered a tie to exist only when the duplication was above what one would expect by random chance. For instance, if in a given time period, a certain website 'A' has a unique user reach of 10% of all Internet users and website 'B' has a reach of 50%, then assuming the consumption of both are independent events, 5% would be the expected number of users to visit both A and B. For all such pairs, we considered a tie only if the observed duplication was greater than the expected value. Hence the value of ties in the network is the *greater than expected duplication* between websites. For some measures we used *dichotomized* ties, where we considered the presence of a tie as '1' and absence as '0.'

Having obtained this network, we first performed some descriptive analysis to ascertain its overall structure. We conducted these analyses as the level of the whole network as well as for individual nodes. We first report some aggregate measures for the whole graph. *Clustering coefficient* (varies between 0 and 1) indicates the average tendency of any three nodes in the graph to form a triangle (i.e. a connected triad). A high value of this measure suggests that a network is composed of communities with high interconnectivity within the communities and low connectivity between them. For this graph, we found clustering coefficient to be quite high (0.846 unweighted and 0.752 weighted, network density =0.395). These suggest that websites cluster into groups in a manner where all sites belonging to the same group have high audience duplication between them and websites belonging to different groups have relatively low duplication between them. A visual inspection of the websites also confirmed the presence of sharp clustering where websites in the same language or those that cater to the same geographies seemed to cluster together (see Figure 2).

**Figure 2 about here**

**Identifying Culturally Defined Markets (CDMs)**

Since this audience duplication network suggested the presence of tightly knit clusters both by visual inspection and by network analytic measures, we performed a hierarchical clustering of the audience duplication matrix. We used the *greater than expected duplication* as the measure of similarity between websites. In other words, the greater the percentage of shared audiences between two websites above the expected duplication, the closer we considered them to be. We obtained a number of solutions as common in hierarchical clustering procedures. Based on the dendogram, we chose the one with 37 clusters, although many of these clusters were single websites or pairs of websites which bridged these clusters. Aside from these bridges, which we explain later, we were left with 953 websites (out of 973 in our sample) that clustered into 18 communities.

These communities conform well to our definition of culturally defined markets, or CDMs, as websites clustering together are either in the same language or those that cater to the same geography, and at times share both language and geography. The largest of these clusters contains all the websites dominated by audiences from the English speaking countries such as United States, UK, and Australia. Even many global websites with content in multiple languages (such as Facebook and Twitter), mainly owing to their large user-base in the US and UK, are a part of this cluster. The second largest cluster is of websites focusing on Chinese-speaking regions, including sites in both traditional and simplified Chinese. Likewise we observed many more such culturally defined clusters, which we list in Table 1 with further information on the languages and/or geographical focuses of their constituent websites.

The other possible solutions (based on the dendogram) also revealed clusters along geo-linguistic lines. A solution with a smaller number of clusters combined sites from Japan, Indonesia and Thailand with those from the UK and US. Another solution with a larger number of clusters resulted in more than 200 clusters. Both these solutions were not conducive to our intended level of analysis, which is to analyze behaviors across countries.

**Table 1 about here**

At first glance, it is tempting to interpret that these clusters are essentially a manifestation of the language of the websites, but language provides only a limited explanation. For instance, our fourth largest cluster, while dominated by French language websites, also has Arabic Websites (from Morocco, Egypt and pan-regional sites such as Arabic Wikipedia and Arabic Yahoo) and hence is more appropriately interpreted as a cluster of websites centered on the Francophone culture. Likewise all India-focused websites, although in English, segregate into a cluster of their own and not with other English language websites that focus on the UK and USA. An exception to our geo-linguistic clusters is a cluster of sites focusing on football (soccer). A further analysis of what explains membership to each cluster is outside the scope of this paper, but our analysis of the Chinese cluster (Cluster No. 2 in Table 1, henceforth referred to as the "C-Cluster") that follows, would shed light on other clusters as well, as we expect that similar factors to explain membership to each CDM

Many of the 20 sites that did not go into any of these clusters are actually sites that bridge these clusters. In network analytic terms, these bridging nodes appear on a large number of shortest paths from any node in the network to any other node. In this website audience duplication network, such websites are accessed by audiences from more than one CDM. Indeed they included content neutral platforms like file sharing platforms, as well as large technology companies such as Windows Media Player, Acrobat, and Intel, whose many language / country specific websites are part of the same domain.

**Relative Isolation of Culturally Defined Markets**

The first impression figure 1 may convey that the WWW is composed of two communities: the China and the World. Access blockage may seem the most obvious reason behind this isolation of the Chinese Websites. However, this CDM is most salient on the visualization simply because China has the largest online population; also 190 websites (out of 973) target China, the largest for any country in our sample. To confirm this intuition, we computed a measure of isolation based on the *group closeness centrality* of each of these CDMs (Everett and Borgatti 1999). This group level measure indicates the average distance of the shortest path between nodes in this CDM, considered as a group, to all other nodes in the network outside of this CDM. For instance, a CDM with sites that shares audiences with all other sites in the network will have a closeness of 1, as it is only step removed from all other nodes. The higher the closeness centrality the more isolated we regard the CDM from the rest of the WWW.

We report the isolation (based on the measure of group closeness centrality) in Table 1. As one would have expected the Global CDM is the least isolated cluster. The Indian CDM is a close second, which is not surprising as the Indian sites included here have English as their primary language. The Chinese CDM is among the clusters with relatively low isolation, and its score is nearly equal to that of the Japanese CDM. In fact most linguistically determined CDMs are more isolated than either the Chinese or Japanese CDMs. In particular the Turkish, Korean, Vietnamese,

Italian and Polish CDMs are the most isolated clusters in that order. In general both language and geographic focus of the websites appear to contribute to the isolation of a CDM.

**C-Cluster as the Chinese CDM on the WWW**

The C-Cluster consists of 194 websites, most of which are in simplified Chinese language, with the remaining handful of sites operating in traditional Chinese. Compared to Chinese-language sites outside, such as the Chinese version of many multinational corporate websites, members in the C-Cluster tend to carry information or services with distinct "Chinese characteristics." For example, music sharing sites feature Chinese artists, social networking sites build around Chinese schools and companies, online stores deliver to Chinese cities, and news portals monitor the ups and downs of Chinese territories. Here a complete spectrum of websites covers all aspects of everyday life, including 33 information portals/search engines, 26 online services such as thematic discussion forums and resource sharing sites, 24 websites providing multimedia entertainment, 20 gaming sites, 18 retail shopping sites, 7 social media sites, and 6 websites run by traditional news media. It is not difficult to imagine Internet users to comfortably inhabit such a self-sufficient virtual world.

An interesting observation from the graph (Figure 1) is the nature of sites within the C-cluster that connect it with the rest of the network. Among such sites the most salient bridge is alibaba.com, a China-based online business-to-business trading platform for small businesses. Boosted by China's manufacturing capacity and being the largest of its kind, it attracts visitors worldwide. More intriguing, however, are the other bridging sites, including the Chinese Wikipedia. Unlike in other CDMs, where Wikipedia tends to have its various language versions occupy the center, its Chinese domain is pushed to the boundary possibly due to a lack of China-originated participation. Also serving as bridges are several Hong Kong and Taiwanese sites, most of which are blocked by the Chinese government, as well as quite a number of mainland China-based entertainment sites, including video/music sharing platforms and online gaming portals.

We further wanted to examine how cultural proximity and access blockage each explain whether a site becomes a part of the C-Cluster. To examine this we classified websites by four levels of cultural proximity. We label the sites in the Chinese (simplified or traditional) as their principal language and focus on China as their main geography as having the highest cultural proximity. Baidu.com and Weibo.com are examples of such sites. Next we label sites with Chinese as their main language and Hong Kong or Taiwan as their focal geographies as those with 'high' cultural proximity. Wretch (a popular blogging platform in Taiwan) and Yahoo Hong Kong are examples of sites in this category. The next category of sites which we label as having 'low' cultural proximity are those that have Chinese as one of their many languages and Greater China as one among the many geographies they cater to. Examples of such sites are 'global' sites BBC and Wordpress.com. The last category, with 'lowest' cultural proximity, comprises of sites that neither focus on China nor are available in the Chinese language.

In Table 2 we report the number of websites that make it to C-Cluster from each of these levels of cultural proximity (in rows). In each of the cells, we also report the percentage of sites that are blocked. The C-Cluster includes both blocked and unblocked sites. Of its 194 sites, 189 focus on mainland China, only 2% of which are blocked, and 4 sites focus on Taiwan and Hong Kong ('high' cultural proximity), 3 of which are blocked. The sites that are culturally distant in general don't group with other sites in the C-Cluster. Among the multilingual websites (that have Chinese language versions) with a global focus, only 2 websites (Xinuhuanet.com and CNTV.com) out of 106 group with the C-Cluster. Of the other 105 websites in this category that do not make it to the C-Cluster, 25 are actually blocked. Finally, none of the least culturally proximate sites are members of the C-Cluster.

Tables 2 about here

## Discussion

Our study questions the widely held assumption that access blockage results in an isolated Chinese Internet from the so-called globalized WWW. We conceptualized the WWW as a network of shared audiences between websites, and find that, rather than one globalized completely connected community, the Internet manifests itself as a collection of many Culturally Defined Markets, China being one of them. A closer analysis of the Chinese CDM suggests that cultural proximity has a greater role than access blockage in shaping online user behavior. In this section, we discuss the implications of our key findings.

First, these CDMs are essentially communities of websites that share either language or geographical focus and often both. The C-Cluster is one such community and is quite similar in its composition of websites to other CDMs such as Japan, South Korea, and Russia. Each CDM is a self-sufficient set with a wide variety of websites to satisfy the diet of typical Internet users. In particular, we find that the C-Cluster is not restricted to mainland Chinese sites, but also includes websites from Hong Kong and Taiwan. This configuration echoes what the scholarship on regional cultural markets refers to as "Greater China," "defined by the [political, cultural, and economic] interactions among its three primary constituent parts" (Chan 2005, 174). [3]

Second, we find that cultural proximity rather than blockage explains the membership to the C-Cluster. As Table 2 reveals, this cluster is largely made up of the most culturally proximate sites (i.e., sites in Chinese focusing on China) and few such sites are blocked. It also includes all Hong Kong and Taiwan focused sites, despite the fact that the GFW blocks most of them. The effect of the GFW is less clear among websites that have Chinese as one of the languages and a "Global" geographic focus. Of them, the two sites that group with the C-Cluster are Chinese state-run news agency and television network, each attempting a global audience. Among the non-members in this category, nearly one fourth are blocked. It is possible that some could have made it into the C-Cluster were it not for the GFW. Blocked sites such as Facebook and YouTube are after all used by the Chinese diaspora. However, several cases suggest that the amount of coercive intervention by the GFW is rather limited. For example, before it finally withdrew in 2010, Google unsuccessfully struggled against Baidu for two years for a share of the Chinese search engine market. Also, although the GFW blockage on the Chinese Wikipedia was lifted in 2008, its users continue to be predominantly from Hong Kong, Taiwan, and the Chinese diaspora. The mainland Chinese have instead stayed with Hudong and Baidu Baike, two local analogues of Wikipedia that emerged during the latter's absence (Ng 2013).

Third, we find that the Chinese CDM is not more isolated than other CDMs. Notably, as indicated by the closeness centrality scores (Table 1), we find that countries allowing much more "open" WWW both in Asia (e.g., Japan) and elsewhere (e.g., Germany and Italy) (Freedom House 2013) constitute CDMs that are just as isolated from the rest of the WWW as the Chinese CDM. In other words, our findings suggest that "the Balkanization of the Internet," as warned against by many (Goldsmith and Wu 2008), is driven primarily by cultural diversity, whose impact is, when applicable, only enhanced by national filtering. Therefore, the prominent narrative about a state-cast "digital curtain" fracturing global flows of Internet communication is highly problematic; Internet blockage like China's GFW should not be the sole focus when assessing a nation's virtual habitat on the WWW.

Our findings then raise the question: Does this mean that, in terms of actual use patterns, the Chinese are not particularly parochial because of the GFW? The three mechanisms that bridge the C-Cluster and the rest of the WWW provide some insight. First, certain relatively cosmopolitan users from China access both domestic and unblocked foreign sites. Second, many users in mainland China visit blocked sites in addition to allowed sites via filter-circumvention techniques. Such practices have their precursors such as picking up of television signal spillovers, music CD smuggling, and film piracy, all of which have long persisted and served to undermine Chinese government's restrictions (Chan 2005; De Kloet 2010; Wu 2012). Third, the audience overlap between websites in and outside the C-Cluster could also be due to users in Hong Kong and Taiwan, as well as the overseas Chinese scattered across the world, who frequent mainland Chinese sites. The fact that the bridging sites either focus on Hong Kong and Taiwan, or are user-generated content sites from mainland China, supports this last explanation. Web usage of the Chinese diaspora, the largest and most widely spread diaspora in the world, may also partially explain the lower isolation of the C-Cluster.

To ascertain how much the behavior of each of these audience segments accounts for the bridging area wherein websites both in and outside the C-Cluster are visited, data at the respondent level is required in order to analyze a two-mode network with both users and websites as nodes. Unfortunately, we do not have access to such data, and hence are unable to confirm to what extent each of aforementioned mechanisms is at work. Regardless, all these mechanisms suggest that even though blockage prevents access to foreign websites, unlike traditional import control, it cannot stop cross-border interactions among ordinary people. These interactions between people in and outside of China generate a new dimension of cultural flows on the WWW.

In sum, although the GFW is a form of geographically focused Internet regulation, our findings show that it does not create a grand partition between the "Chinternet" from the global Internet. Further, our results challenge the predominant discourse in Western policy-making, popular media, and oftentimes academic works, which assume the sole function of access blockage to be coercive interference in user behavior.

Based on these findings, it is worth speculating how the lifting of blockage would reshape the Chinese CDM. First, it is most likely that the existing websites in the C-Cluster would continue to stick together due to cultural proximity. The major changes would be in the bridging area, which contains currently blocked websites from Hong Kong and Taiwan. If these become freely accessible to the mainland Chinese, it is reasonable to expect that they would be pulled closer to the densely connected mainland sites, as they carry content appealing to China-based users. However, we contend that these sites would not become central within the cluster, as they have a strong local focus on Hong Kong and Taiwan, and many mainland Chinese may find either their ideological overtones unpleasant or their content irrelevant. Even for blocked websites with most culturally proximate content, lifting the blockage would not lead to dramatic changes in their locations. For example, as already noted, after four years of blockage and made accessible again in 2008, the Chinese Wikipedia has failed to accumulate a significant user base in China (Ng 2013). We also expect that the C-Cluster as a whole would move closer to the rest of the WWW, as more Chinese would participate in global social media sites such as Facebook, Twitter, and YouTube. Nonetheless, a distinct Chinese CDM would remain as do its Japanese, Russian, and South Korean counterparts, because users consume culturally proximate content, with or without access blockage. Therefore, compared to removing the GFW of China, on which most policy, popular, and scholarly discourse tend to concentrate, battling against content censorship over domestic websites may bring about much more substantial changes in what Chinese people make use of on the Internet, which may enable further cultural changes.

Finally, we emphasize a sociological approach when examining government control over Internet use in the form of access blockage. Such an approach departs from normative prescription ingrained in classical liberalism (Cramer 2013). It regards Internet use as cultural consumption situated in specific social conditions. Research on China's film import controls has seen a similar shift where controls were first conceived as ideological maneuvers, then also as cultural protectionism, and finally as one factor amid many in the complex dynamics of Chinese pirate film consumption (Wu 2012). Studies on state censorship in other nondemocratic contexts also advocate empirical interrogations (e.g., Abramitzky and Sin 2010). Our study extends this stance. We demonstrate that Internet blockage should not be taken readily as external coercion against certain hypothetically autonomous and normatively anticipated behavior; instead, it should be investigated as one of many structural aspects that shape media choices.

---

[1] This information has been taken from comScore's own documentation on methodology that is only available to subscribers.

[2] The methods GreatFire uses for detecting GFW blockage are explained at https://en.greatfire.org/faq.

[3] To clarify, our approach aims to describe and analyze existing CDMs, while refraining from reifying a unified cultural logic behind identified CDMs. For instance, by illuminating the existence of a Chinese CDM, we do not imply about "Chineseness" or a possible realization of a "Chinese civilization-state" (Cf. Tu, 2005 [1991]).

# References

Abramitzky, R., and I., Sin. 2010. Book Translations As Idea Flows: The Effects of the Collapse of Communism on the Diffusion of Knowledge (Discussion Paper No. 09-032). Stanford, CA: Stanford Institute for Economic Policy Research, Stanford University.

Bamman, D., O. C., Brendan, and N. A. Smith. 2012. Censorship and deletion practices in Chinese social media. *First Monday* 17: online.  Available at: http://www.uic.edu/htbin/cgiwrap/bin/ojs/index.php/fm/article/view/3943/3169 (accessed March 22, 2013)

Barmé, G. R., and S. Ye. 1997. The Great Firewall of China. *Wired* 5.06: 138-151.

Barnett, G. A., C. J. O. O. Chung, and H. W. Park. 2011. Uncovering Transnational Hyperlink Patterns and Web-Mediated Contents: A New Approach Based on Cracking.com Domain. *Social Science Computer Review* 29: 369–384.

Baughn, C. C., and M. A. Buchanan. 2001. Cultural protectionism. *Business Horizons* 44: 5-15.

Burri, M. 2012. Cultural protectionism 2.0: Updating cultural policy tools for the digital age. In *Transnational culture in the internet age*, ed. A. Candeub and S. Pager, pp. 182-202. Cheltenham, UK: Edward Elgar Publishing.

Cannici, W. J., Jr. 2009. The Global Online Freedom Act: combating American businesses that facilitate Internet censorship in China. *Journal of Internet Law* 12: 3-17.

Chan, J. M. 2005. Trans-border broadcasters and TV regionalization in Greater China: Processes and strategies. In *Transnational Television Worldwide: Towards a New Media Order,* ed. J. K. Chalaby, pp. 173-195. London: I.B.Tauris.

Chao, L., and B. Worthen. 2010. Google Is Poised to Close China Site, *The Wall Street Journal*, , March 15, online.  Available at:

http://online.wsj.com/article/SB10001424052748703457104575121613604741940.html (accessed March 25, 2013)

CNNIC. n.d. Statistical Reports on the Internet Development in China. Available at: http://www1.cnnic.cn/en/index/0O/02/index.htm (accessed March 4, 2013)

Cramer, B. W. 2013. The two Internet Freedoms: Framing victimhood for political gain. *International Journal of Communication* 7: 1074-1092.

Curtin, M. 2003. Media Capital: Towards the Study of Spatial Flows. *International Journal of Cultural Studies* 6: 202–228.

Damm, J. 2007. The internet and the fragmentation of Chinese society. *Critical Asian Studies* 39: 273-294.

Davis, L. L. 2003. Cultural proximity on the air in Ecuador: National, regional television outperforms imported US programming. In *The impact of international television: A paradigm shift*, ed. M. G. Elasmar, pp. 111-131. Mahwah, NJ: Lawrence Earlbaum Associates.

Faris, R., and N. Villeneuve. 2008. Measuring Global Internet filtering. In *Access denied,* ed. R. Deibert, J. Palfrey, R. Rohozinski, and J. Zittrain, pp. 5-27. Cambridge, MA: MIT Press.

Feng, G. C., and S. Z. Guo. 2013. Tracing the route of China's Internet censorship: An empirical study. *Telematics and Informatics*, 30(4): 335-345.

Freedom House. 2013. Freedom on the Net 2012. Available: http://www.freedomhouse.org/report/freedom-net/freedom-net-2012 (accessed March 21, 2013)

Goldsmith, J., and T. Wu. 2008. *Who Controls the Internet?: Illusions of a Borderless World*. New York: Oxford University Press.

Guo, S., and G. Feng. 2011. Understanding support for Internet censorship in China: An elaboration of the theory of reasoned action. *Journal of Chinese Political Science* 17: 33-52.

Harwit, E., and D. Clark. 2001. Shaping the internet in China: Evolution of political control over network infrastructure and content. *Asian Survey* 41: 377-408.

Hassid, J. 2012. Safety Valve or Pressure Cooker? Blogs in Chinese Political Life. *Journal of Communication* 62: 212-230.

Hecht, B., and D. Gergle. 2010. The tower of Babel meets web 2.0: user-generated content and its applications in a multilingual context. In *Proceedings of the 28th international conference on Human factors in computing systems (CHI '10)*, pp. 291–300. New York: ACM

Internet Society. 2013. Global Internet User Survey 2012. Available at: http://www.internetsociety.org/survey (accessed March 11, 2013)

Jacobs, K. 2012. *People's pornography: Sex and surveillance on the Chinese Internet*. Bristol, UK: Intellect.

Jin, D. Y. 2007. Reinterpretation of cultural imperialism: emerging domestic market vs continuing US dominance. *Media, Culture & Society* 29: 753–771.

King, G., J. Pan, and M. Roberts. 2013. How censorship in China allows government criticism but silences collective expression. *American Political Science Review*, 107(2): 326-343.

De Kloet, J. 2010. *China with a Cut: Globalisation, Urban Youth, and Popular Music*. Amsterdam: Amsterdam University Press.

Everett, M. G., and S. P. Borgatti. 1999. The centrality of groups and classes. *The Journal of Mathematical Sociology* 23: 181-201.

Kraidy, M. M. 1999. The global, the local, and the hybrid: A native ethnography of glocalization. *Critical Studies in Media Communication* 16: 456–476.

Ksiazek, T. B., and J. G. Webster. 2008. Cultural Proximity and Audience Behavior: The Role of Language in Patterns of Polarization and Multicultural Fluency. *Journal of Broadcasting & Electronic Media* 52: 485–503.

La Pastina, A., and J. Straubhaar. 2005. Multiple Proximities between Television Genres and Audiences: The Schism between Telenovelas' Global Distribution and Local Consumption. *International Communication Gazette* 67: 271–288.

Li, Y. 2009. *Our Great Firewall: Expression and governance in the era of the Internet* [in Chinese]. Guilin, People's Republic of China: Guangxi Normal University Press.

Liu, C. 2010. Internet Censorship as a Trade Barrier: A Look at the WTO Consistency of the Great Firewall in the Wake of the China-Google Dispute. *Georgetown Journal of International Law* 42: 1199-1237.

Liu, F. 2010. *Urban Youth in China: Modernity, the Internet and the Self*. New York: Routledge.

MacKinnon, R. 2008. Flatter world and thicker walls? Blogs, censorship and civic discourse in China. *Public Choice* 134: 31-46.

MacKinnon, R. 2009. China's Censorship 2.0: How companies censor bloggers. *First Monday, 14*: online. Available: http://firstmonday.org/article/view/2378/2089 (accessed March 10, 2013)

MacKinnon, R. 2011. China's "networked authoritarianism". *Journal of Democracy* 22: 32-46.

Marolt, P. 2011. Grassroots agency in a civil sphere? Rethinking Internet control in China. In *Online Society in China*, eds. D. K. Herold and P. Marolt, pp. 53-67. London: Routledge.

Meng, B. 2011. From Steamed Bun to Grass Mud Horse: E Gao as alternative political discourse on the Chinese Internet. *Global Media and Communication* 7: 33-51.

Mueller, M. L. 2011. China and Global Internet Governance: A Tiger by the Tail. In *Access Contested: Security, Identity, and Resistance in Asian Cyberspace*, eds. R. Deibert, J. Palfrey, R. Rohozinski and J. Zittrain, pp. 177-194. Cambridge, MA: MIT Press.

Mulvenon, J. C., and M. S. Chase. 2005. Breaching the Great Firewall: External challenges to China's Internet control. *Journal of E-Government* 2: 73-84.
29

Neyazi, T. A. 2010. Cultural Imperialism or Vernacular Modernity. *Media, Culture & Society* 32: 907-924.

Ng, J. 2013. The difficulties of identifying censorship in an environment with distributed oversight: a large-scale comparison of Wikipedia China with Hudong and Baidu Baike. Available at: https://citizenlab.org/2013/08/a-large-scale-comparison-of-wikipedia-china-with-hudong-and-baidu-baike/ (accessed October 11, 2013)

Nordenstreng, K. 2011. Free flow doctrine in global media policy. In *The handbook of global media and communication policy*, eds. R. Mansell and M. Raboy, pp. 79–94. Malden, MA: Blackwell.

Pool, I. 1977. The changing flow of television. *Journal of Communication* 27: 139–149.

Schiller, H. I. 1969. *Mass communications and American empire*. New York: A.M. Kelley.

Straubhaar, J. 1991. Beyond media imperialism: Assymetrical interdependence and cultural proximity. *Critical Studies in Media Communication* 8: 39–59.

Straubhaar, J. 2003. Choosing national TV: Cultural capital, language, and cultural proximity in Brazil. In *Impact of International Television: A paradigm shift*, ed. M. G. Elasmar, pp. 77–110. Mahwah, NJ: Lawrence Erlbaum.

Straubhaar, J. 2007. *World television : From global to local*. Los Angeles: Sage.

Takhteyev, Y., A. Gruzd, and B. Wellman. 2012. Geography of Twitter networks. *Social Networks* 34: 73–81.

Taneja, H., J. G. Webster, E. C. Malthouse, and T. B. Ksiazek. 2012. Media consumption across platforms: Identifying user-defined repertoires. *New Media & Society* 14: 951–968.

Tsui, L. 2008. The Great Firewall as Iron Curtain 2.0: The implications of China's Internet most dominant metaphor for U.S. foreign policy. Paper presented at the The 6th annual Chinese Internet Research Conference, Hong Kong, June 13-14.

Tu, W.-m. 2005 [1991]. Cultural China: The Periphery as the Center. *Daedalus* 134: 145-167.

Webster, J. G. 1985. Program audience duplication: A study of television inheritance effects. *Journal of Broadcasting & Electronic Media* 29: 121–133.

Webster, J. G., and T. B. Ksiazek. 2012. The Dynamics of Audience Fragmentation: Public Attention in an Age of Digital Media. *Journal of Communication* 62: 39–56.

Wu, A. X. 2012. Broadening the scope of cultural preferences: Movie talk and Chinese pirate film consumption, 1980s - 2005. *International Journal of Communication* 6: 501-529.

Xiao, Q. 2011. The Battle for the Chinese Internet. *Journal of Democracy* 22: 47-61.

Yang, G. 2009. *The power of the Internet in China: Citizen activism online*. New York: Columbia University Press.

Zhang, L. L. 2006. Behind the 'Great Firewall': Decoding China's Internet Media Policies from the Inside. *Convergence* 12: 271-291.

Zheng, Y. 2007. *Technological Empowerment: The Internet, State, and Society in China*. Stanford, CA: Stanford University Press.

**Figure 1 Internet Access Blockage and User Behavior**

a. Extant view: User choice based on access

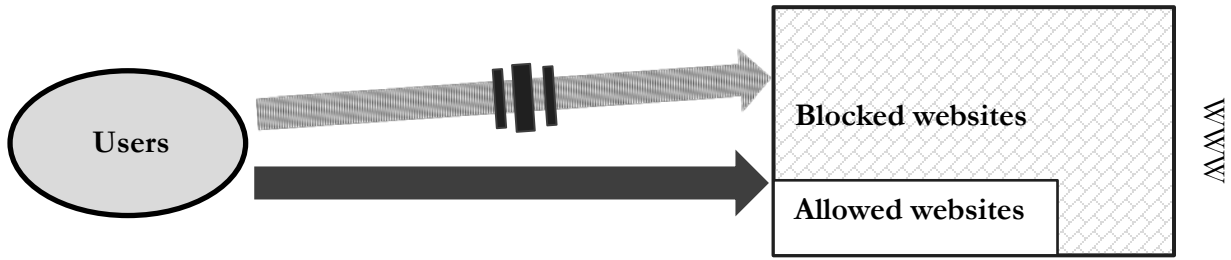

b. Proposed view: User choice based on cultural proximity *

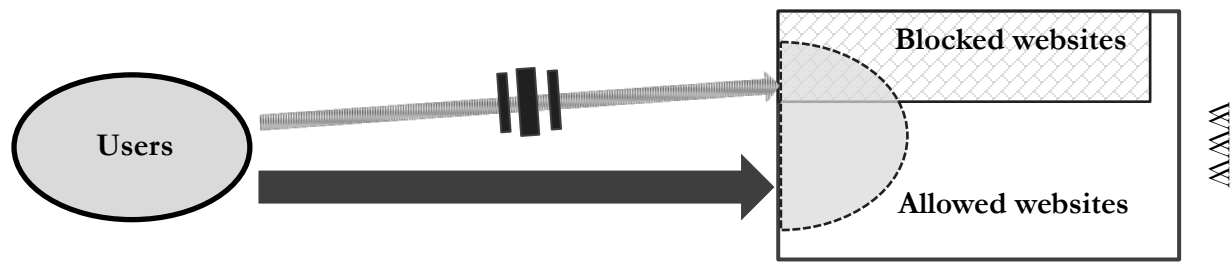

b1. t = around 2001: imposition of blockage

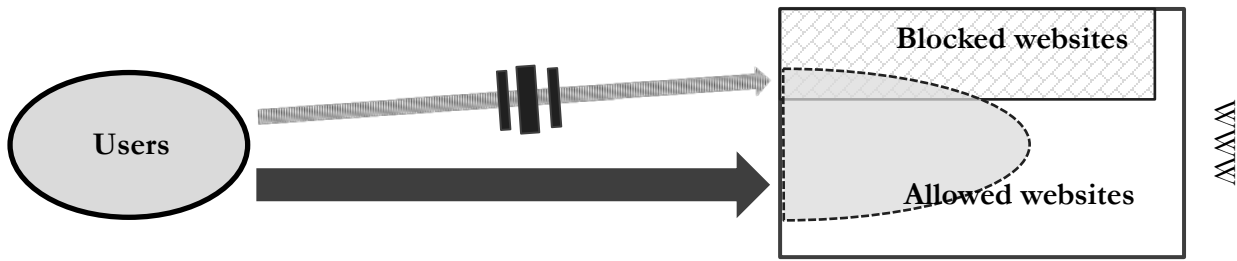

b2. t = 2012: after years of persistent blockage

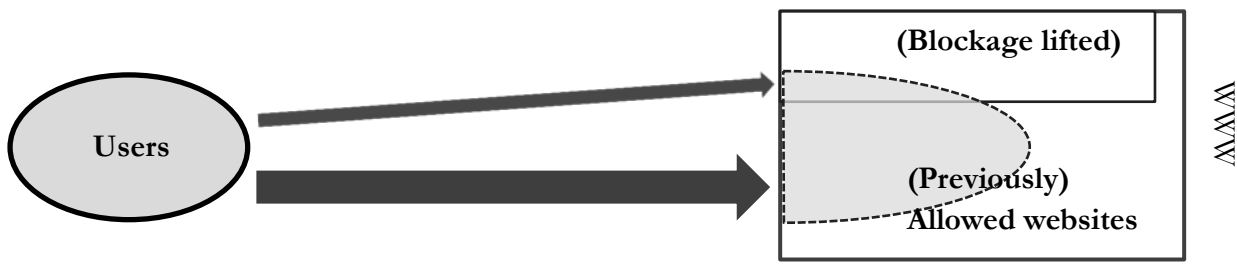

b3. What if blockage is lifted today?

➤ Enacted user preferences    ▐▐▐▷ Unfulfilled user preferences    ⌒ Culturally proximate content

* Blocked websites are far less in proportion than commonly implied in the dominant discourse. See our data analyses that follow.

**Figure 2 Visualization of Culturally Defined Markets[3]**

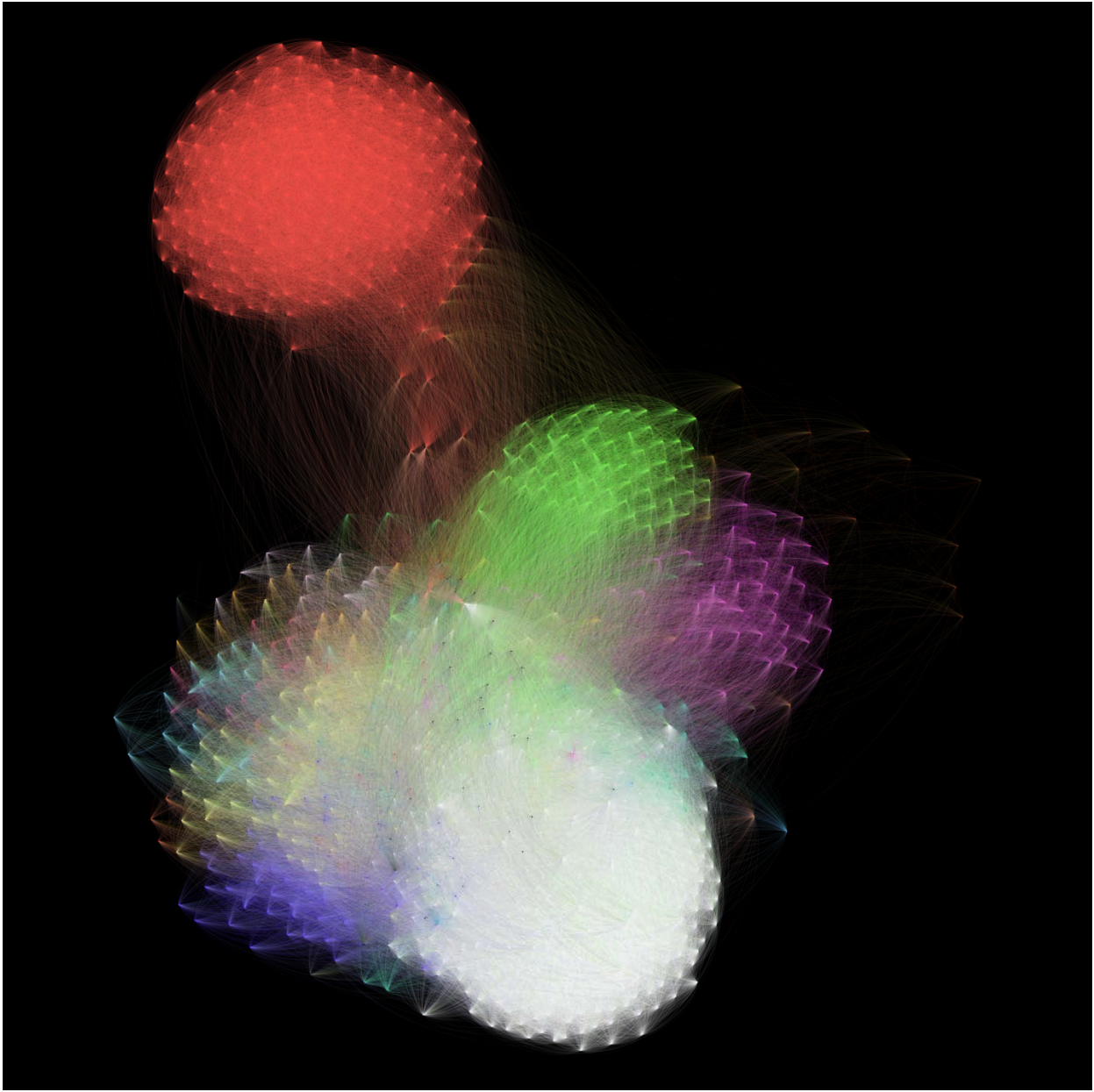

---

[3] Each color corresponds to a culturally defined market. See Table 1 for legend.
NOTE: If you are reading a black-and-white version, please access the colored version of both Figure 2 and Table 1 at: http://angelaxiaowu.com/files/graph.png.

**Table 1 Culturally Defined Markets Explained (Cluster Solution)**

| Cluster No. | No of Sites | Language | Geography | Other Characteristics | Isolation |
|---|---|---|---|---|---|
| 1 | 351 | English | Global, US, UK, Canada, Australia | | 1.05 |
| C-Cluster; 2 | 194 | Chinese | China, HK, Taiwan | | 1.13 |
| 3 | 111 | Japanese | Japan, Global | File-sharing | 1.14 |
| 4 | 61 | French, Arabic | France, the Middle East | | 1.19 |
| 5 | 47 | Spanish | Spain, Argentina, Mexico | | 1.21 |
| 6 | 47 | Russia | Russian | | 1.22 |
| 7 | 22 | Portuguese | Brazil, Portugal | | 1.21 |
| 8 | 21 | German | Germany | | 1.26 |
| 9 | 19 | English | India | | 1.09 |
| 10 | 16 | Polish, Multiple | Poland, Global | Sports betting[4] | 1.32 |
| 11 | 15 | Korean | South Korea | | 1.49 |
| 12 | 10 | Italian | Italy | | 1.43 |
| 13 | 10 | Turkish | Turkey | | 1.72 |
| 14 | 8 | English | Africa, Global | Social networking[5] | 1.27 |
| 15 | 6 | Vietnamese | Vietnam | | 1.43 |
| 16 | 5 | Bahasa, Thai | Indonesia, Thailand | | 1.31 |
| 17 | 5 | Multiple | Global | Soccer / Football | 1.21 |
| 18 | 3 | Dutch | The Netherlands | | 1.40 |

---

[4] Sports betting is legal in Poland.
[5] These are second tier (excluding Facebook, Twitter, and LinkedIn) social networking sites and have a disproportionately large number of users in African countries. This could be because they need lower bandwidth than most others.
NOTE: If you are reading a black-and-white version, please access the colored version of both Figure 2 and Table 1 at: http://angelaxiaowu.com/files/graph.png.

**Table 2 C-Cluster Membership**

| Classification of Website | Cultural Proximity | Chinese Cluster | |
| --- | --- | --- | --- |
| | | Non-Member | Member |
| In Chinese* and China Focus | Highest | 0 *(0%)* | 189 *(2%)* |
| In Chinese and Taiwan & HK Focused | High | 0 | 4 *(75%)* |
| In Chinese but Global Focus | Low | 105 *(24%)* | 1 *(0%)* |
| Neither in Chinese nor Greater China Focused | Lowest | 674 *(10%)* | 0 |
| Total Number of Websites | | 779 *(12%)* | 194 *(4%)* |

No. of Websites *(% of these websites blocked from access in China)*

* "In Chinese" refers to websites that offer content in the Chinese language, simplified or traditional.